\documentclass[sigconf]{acmart}
\AtBeginDocument{%
  \providecommand\BibTeX{{%
    \normalfont B\kern-0.5em{\scshape i\kern-0.25em b}\kern-0.8em\TeX}}}

\setcopyright{none}

\usepackage{multirow}
\usepackage{mdframed}
\usepackage{framed,enumitem} 
\usepackage{graphicx}
\usepackage{xspace}
\usepackage{makecell}
\usepackage{soul}
\usepackage{xcolor}
\usepackage{textcomp}
\usepackage{array}
\usepackage{longtable}
\usepackage{tabularx}
\usepackage{listings}
\usepackage{subcaption} 

\newcommand{\sysname}{\textsc{STEPwise}}
\newcommand{\fname}{\textsc{STEP}}
\newcommand{\numTeachers}{5}
\newcommand{\numStudents}{64}

\definecolor{redbox}{HTML}{FF8B78}
\definecolor{orangebox}{HTML}{FF9C45}
\definecolor{greenbox}{HTML}{529952}
\definecolor{bluebox}{HTML}{6CA2FF}
\definecolor{purplebox}{HTML}{7D67BC}
\definecolor{pinkbox}{HTML}{FF6AAF}

\newcommand{\redbox}[1]{\colorbox{redbox}{\textcolor{white}{\textbf{#1}}}}
\newcommand{\orangebox}[1]{\colorbox{orangebox}{\textcolor{white}{\textbf{#1}}}}
\newcommand{\greenbox}[1]{\colorbox{greenbox}{\textcolor{white}{\textbf{#1}}}}
\newcommand{\bluebox}[1]{\colorbox{bluebox}{\textcolor{white}{\textbf{#1}}}}
\newcommand{\purplebox}[1]{\colorbox{purplebox}{\textcolor{white}{\textbf{#1}}}}
\newcommand{\pinkbox}[1]{\colorbox{pinkbox}{\textcolor{white}{\textbf{#1}}}}

\definecolor{blockcolor}{HTML}{000000}
\definecolor{blockrule}{gray}{0.6}
\newmdenv[
  topline=false,
  bottomline=false,
  rightline=false,
  leftline=true,
  linecolor=blockrule,
  linewidth=1pt,
  innertopmargin=2pt,
  innerbottommargin=2pt,
  innerleftmargin=10pt,
  innerrightmargin=10pt,
  skipabove=10pt,
  skipbelow=10pt,
  backgroundcolor=white,
  font=\small\color{blockcolor}
]{block}

\newmdenv[
  topline=true,
  bottomline=true,
  rightline=true,
  leftline=true,
  linecolor=blockrule,
  linewidth=1pt,
  innertopmargin=5pt,
  innerbottommargin=5pt,
  innerleftmargin=5pt,
  innerrightmargin=5pt,
  skipabove=5pt,
  skipbelow=10pt,
  backgroundcolor=white,
  font=\sffamily\color{blockcolor},
]{task}

\lstnewenvironment{prompt}{%
  \lstset{
    breaklines=true,
    linewidth=\linewidth,
    columns=fullflexible,
    keepspaces=true,
    basicstyle=\ttfamily\footnotesize,
    frame=single
  }
}{}

\begin{document}


% \title{Diagnosing Problem-Solving Skills through LLM-Generated Stepwise Sub-Questions}
\title[STEPwise: Diagnosing Mathematical Problem-Solving Skills with LLM-Generated Questions and Rubrics]{\sysname{}: Supporting Teachers in Diagnosing Students’ Mathematical Problem-Solving Skills through LLM-Generated Questions and Rubrics}

\author{Yoonsu Kim}
\orcid{0000-0002-9782-086X}
\affiliation{\institution{School of Computing \\ KAIST}
\city{Daejeon}
\country{Republic of Korea}}
\email{yoonsu16@kaist.ac.kr}

\author{Hayeon Doh}
\affiliation{\institution{Department of AI Computing \\ KAIST}
\city{Daejeon}
\country{Republic of Korea}}
\email{hannahdoh@kaist.ac.kr}

\author{Eunhye Kim}
\affiliation{\institution{School of Computing \\ KAIST}
\city{Daejeon}
\country{Republic of Korea}}
\email{gracekim027@kaist.ac.kr}

\author{Jinho Son}
\affiliation{\institution{Algorithm Labs}
\city{Seoul}
\country{Republic of Korea}}
\email{sjhfam@algorithmlabs.co.kr}

\author{Juho Kim}
\email{juhokim@kaist.ac.kr}
\orcid{0000-0001-6348-4127}
\affiliation{
    \institution{School of Computing, KAIST}
    \city{Daejeon}
    \country{Republic of Korea}
}
\email{juho@skillbench.com}
\affiliation{
    \institution{SkillBench}
    \city{Santa Barbara, CA}
    \country{USA}
}

\renewcommand{\shortauthors}{Yoonsu Kim et al.}




% %% A "teaser" image appears between the author and affiliation
% %% information and the body of the document, and typically spans the
% TEASER FIGURE
\begin{teaserfigure}
\centering
  \includegraphics[width=\textwidth]{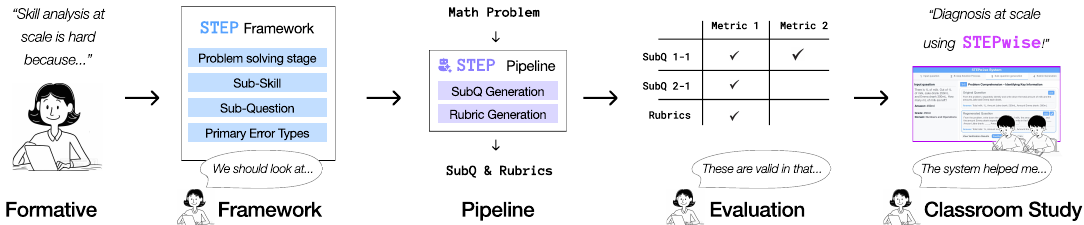}
  \caption{Overview of the research. We conduct a formative study with teachers on what makes problem-solving skill diagnosis at scale difficult. Based on the results, we construct \fname{}, which consists of 4 problem-solving stages with 8 sub-skills. Based on \fname{}, we create an LLM-based pipeline that generates sub-questions and rubrics from math word problems, and evaluate them with teachers. Finally, we present \sysname{}, a system that incorporates the LLM-pipeline based on \fname{} and supports teachers in diagnosing students' problem-solving skills, and examine its classroom use through a month-long study with \numTeachers{} teachers and \numStudents{} students.}
  \Description{A five-stage overview diagram showing the research process from left to right, connected by arrows. Stage 1 (Formative): An illustration of a teacher with a speech bubble saying "skill analysis at scale is hard because..." Stage 2 (Framework): The STEP Framework box listing four components -- Problem solving stage, Sub-Skill, Sub-Question, and Primary Error Types -- with a speech bubble "We should look at..." Stage 3 (Pipeline): A math problem flows into the STEP Pipeline, which contains SubQ Generation and Rubric Generation modules, producing sub-questions and rubrics as output. Stage 4 (Evaluation): A table showing SubQ 1-1, SubQ 2-1, and Rubrics evaluated against Metric 1 and Metric 2, all marked with checkmarks, with a speech bubble "These are valid in that..." Stage 5 (Classroom Study): An illustration of a teacher reading student work alongside a screenshot of the STEPwise system interface, with a speech bubble "The system helped me..." and the text "Diagnosis at scale using STEPwise!"}
\label{fig:teaser}
\end{teaserfigure}

\begin{abstract}
Problem-solving is central to mathematics education. However, diagnosing where students’ reasoning breaks down requires individualized probing and consistent criteria that teachers cannot sustain at classroom scale. We introduce \fname{} (STage-based Evaluation of Problem-solving skills), co-designed with six mathematics education experts, which operationalizes problem-solving as four stages and eight diagnostic sub-skills. By specifying what evidence to elicit and assess at each stage, \fname{} enables an LLM pipeline to generate stage-specific questions and rubrics for any math word problem, reducing the manual design of diagnostic materials. An expert review (N=8) assessed the pedagogical validity of the generated materials and informed \sysname{}, a teacher-in-the-loop system for diagnostic material generation and process-level assessment. In a classroom study with \numTeachers{} teachers and \numStudents{} students, \sysname{} helped teachers and students locate reasoning breakdowns and supported students in reflecting on their reasoning. Together, \fname{} and \sysname{} offer a structured approach to fine-grained diagnosis at classroom scale.

% Problem-solving is central to mathematics education. However, fine-grained diagnosis requires teachers to design targeted probes and interpret each student’s reasoning consistently—work difficult to repeat across problems and students. We introduce \fname{} (STage-based Evaluation of Problem-solving skills), co-designed with six mathematics education experts, which operationalizes problem-solving as four stages and eight observable sub-skills. By specifying what evidence to elicit and assess at each stage, \fname{} provides a reusable schema through which an LLM pipeline generates targeted questions and rubrics for math word problems, reducing this manual design bottleneck. An expert review (N=8) assessed the pedagogical validity of the generated materials and motivated \sysname{}, a teacher-in-the-loop system for reviewing materials and diagnosing responses. In a classroom study with \numTeachers{} teachers and \numStudents{} students, \sysname{} helped teachers and students locate reasoning breakdowns and scaffolded reflection. These findings demonstrate a practical approach to scaling process-level diagnosis while retaining teacher oversight.
\end{abstract}
\maketitle
\section{Introduction}
Problem-solving is central to mathematics education. Successfully solving a problem requires students not only to know mathematical concepts, but also to interpret the problem situation, identify relevant knowledge, and organize and carry out a solution. Accordingly, mathematics educators regard students’ problem-solving processes as important instructional targets~\cite{anderson2009mathematics, ferrini2000principles, lee2016appropriate}.

Diagnosing students' performance at each stage of the problem-solving process is essential for supporting their learning~\cite{glaser2001knowing, schoenfeld2007assessing}. Yet in practice, it remains heavily dependent on teachers' professional judgment, grounded in their close examination of students' problem-solving behavior or written work~\cite{schoenfeld2014mathematical, philipp2014diagnostic, corno2008teaching}. Since students do not consistently externalize their reasoning without explicit support~\cite{renkl2014toward, chi1989self}, prior work has emphasized designing tasks that intentionally elicit students' thinking~\cite{kelly2016analysing, lee2016appropriate}, such as written explanation prompts. 
In parallel, research on mathematical problem solving has proposed stage-based frameworks that decompose the solving process into structured phases~\cite{polya2004solve, schoenfeld1983beyond, newman1977analysis, jin2024using}, offering conceptual guidance on what kinds of reasoning may occur at each step.
However, it remains unclear how these general stages can be translated into problem-specific questions that elicit diagnostically informative responses. It also remains unclear how teachers can consistently assess those responses within everyday classroom workflows.

Motivated by this gap, we conducted a formative study with four mathematics teachers (mean teaching experience: 13.25 years) and 48 students in grades 3--6. Using four general stage-based questions adapted from prior work~\cite{jin2024using}, we examined how teachers diagnosed students’ problem-solving processes and what challenges they encountered. We identified two challenges. First, questions that were not tailored to individual problems often failed to elicit the reasoning needed to solve them. Second, teachers needed clear, problem-specific rubrics to assess students’ reasoning consistently, but constructing such rubrics demanded substantial time and pedagogical effort. These challenges pointed to the need for a framework that specifies both what reasoning to elicit and what evidence to assess at each stage.

To address this need, we developed \fname{} (STage-based Evaluation of Problem-solving skills), a diagnostic framework co-designed with six mathematics education experts and informed by established problem-solving frameworks~\cite{polya2004solve, schoenfeld1983beyond, krulik1988problem, hashemi2015designing, thiangthung2016applying}. \fname{} organizes problem solving into four sequential stages and refines each stage into two diagnostic sub-skills. These sub-skills distinguish sources of difficulty within a broad stage—for example, identifying key information versus identifying the problem goal, or applying mathematical knowledge versus constructing a mathematical expression. By specifying what evidence to elicit and assess for each sub-skill, \fname{} helps locate where students’ reasoning begins to break down.

Building on \fname{}, we developed an LLM-based pipeline that instantiates this diagnostic structure for individual math word problems. Given a problem, the pipeline generates a sub-question targeting each sub-skill and a corresponding rubric that characterizes different levels of student reasoning. Automating this problem-specific generation reduces the manual effort required to create diagnostic materials across problems, while the shared structure supports consistent process-level assessment.

We evaluated our approach in two stages. First, eight mathematics teachers (mean teaching experience: 11.83 years) participated in an expert review of materials generated for 36 problems spanning grades 3--6. Six conducted the primary review, while two additional teachers resolved disagreements. At the sub-skill level, 91.32\% of the materials met the construct-validity criterion, while 90.97\% met the mathematical-accuracy criterion. However, level appropriateness (80.21\%) and problem-context alignment (72.22\%) received lower ratings, revealing occasional mismatches with grade-level language, curricula, and problem contexts. These limitations made teacher review and adaptation a central design requirement. Accordingly, we incorporated teachers’ curriculum materials into the generation pipeline and enabled teachers to review and edit its outputs. These refinements informed \sysname{}, a teacher-in-the-loop system for diagnostic material generation and process-level assessment.

We then conducted a month-long classroom study with \numTeachers{} teachers and \numStudents{} students in grades 3--9 to examine how \fname{} and \sysname{} support process-level diagnosis in classroom use. Over the month, teachers generated diagnostic materials for 90 problems in total, yielding 720 sub-questions and corresponding rubrics.
Teachers reported two main benefits: they could produce diagnostic materials for their own problems easily, and the step-by-step responses let them see how far students had progressed and pinpoint where the problem-solving process began to break down---a level of detail they could otherwise obtain only through individual follow-up questioning.
They further noted that beyond diagnosis, the sub-questions themselves serve as a learning experience, guiding students to work through one problem thoroughly and helping them develop structured thinking beyond simply applying memorized procedures.
Students responded positively to the stage-based approach in terms of its usefulness for working through problems, expressing their reasoning, recognizing their strengths and weaknesses, and willingness to use it for future learning (all means between 3.6 and 3.9 out of 5; \autoref{fig:classroom_likert}).
At the same time, the format was not equally easy for everyone: the students who struggled most with writing out each stage were those least accustomed to articulating their reasoning. Teachers noted that these students are the ones who need this kind of diagnosis and practice most, but suggested that for them adjacent stages could be merged into fewer sub-questions to reduce their burden.

This work makes three contributions. First, we introduce \fname{}, a diagnostic framework co-designed with mathematics education experts that refines four problem-solving stages into eight sub-skills and specifies what evidence to elicit and assess at each stage. Second, we present an LLM-based pipeline that generates problem-specific sub-questions and rubrics based on \fname{}, and \sysname{}, a teacher-in-the-loop system that supports diagnostic material generation and process-level assessment. Third, through a formative study, an expert review, and a month-long classroom deployment, we provide empirical findings on the practical challenges of fine-grained diagnosis, the strengths and limitations of LLM-generated diagnostic materials, and how teachers and students engage with stage-based diagnosis in classroom practice.

% This work makes three contributions. First, we introduce \fname{}, a structured problem-solving diagnostic framework for math word-problem solving, grounded in prior literature and co-designed with mathematics education experts. 
% Second, we operationalize this framework at scale through an LLM-based pipeline that automatically generates stage-specific sub-questions and rubrics for a given problem, and \sysname{}, a teacher-in-the-loop system spanning material generation to student assessment.
% Third, through an expert review and a month-long classroom deployment with \numTeachers{} teachers and \numStudents{} students, we provide evidence for the mathematical validity and classroom value of process-level diagnosis, along with findings on how its granularity should adapt to problem difficulty and student readiness.

\section{Related Work}
Our work builds on three interconnected areas: theoretical frameworks of mathematical problem-solving stages, practical approaches to assessing and supporting these processes in educational settings, and recent advances in using AI for education. 

\subsection{Problem-Solving Process Frameworks in Mathematics Education}
Understanding and teaching problem-solving as a process has been a central concern in mathematics education for decades. Polya's seminal work established a foundational four-stage model consisting of understanding the problem, devising a plan, carrying out the plan, and looking back~\cite{polya2004solve}, emphasizing mathematical problem-solving as deliberate cognitive activities at each stage~\cite{polya2004solve}. Studies have shown that Polya's framework leads to significantly greater improvements in students’ word-problem performance~\cite{fitriani2025effectiveness, gulam2024mathematics}. Building on Polya's work, Schoenfeld extended the framework by incorporating metacognitive dimensions~\cite{schoenfeld1983beyond, schoenfeld2014mathematical}. Schoenfeld argued that successful problem-solving requires not only domain knowledge and strategies but also metacognitive control---the ability to monitor one's progress, evaluate the effectiveness of strategies, and make adaptive decisions during the solving process. They highlighted that students often possess relevant knowledge but fail to deploy it effectively due to poor metacognitive regulation. 

Newman's error analysis~\cite{newman1977analysis} provided another influential perspective by identifying specific types of errors students make at different stages of problem-solving. Newman's hierarchy distinguishes between reading errors, comprehension errors, transformation errors, process errors, and encoding errors. This framework has been particularly useful for diagnostic purposes, as it allows teachers to pinpoint where in the problem-solving process students encounter difficulties~\cite{rohmah2018Newman, sambo2025using, lupita2025analysis}. Other researchers have proposed variations and extensions of these frameworks. For example, Krulik and Rudnick further specified the set of heuristics that are successful at all levels of problem-solving~\cite{krulik1988problem}. Goulet-Lyle et al. highlight the need to establish links between the explicit problem content and students’ knowledge while devising a plan~\cite{goulet2020does}.

Despite the richness of these theoretical frameworks, contemporary mathematics education faces new challenges. Research has highlighted that students struggle with the early stages of problem-solving, particularly in comprehending textual information in word problems, due to a growing tendency toward passive media consumption~\cite{singer2017reading, jin2024using}.
These evolving challenges highlight the need for practical approaches that can guide and assess students' problem-solving processes.

\subsection{Assessing and Supporting Problem-Solving Processes: Challenges and Approaches}
Assessing students' problem-solving processes requires more than evaluating their final answers. Researchers have argued that effective diagnosis must capture how students think and approach across each stage of problem-solving~\cite{glaser2001knowing, schoenfeld2007assessing}. However, making these cognitive processes visible has proven challenging, as much of students' reasoning remains implicit unless explicitly externalized~\cite{chi1989self, henderson2004grading}. 
Several approaches have been developed to elicit students' thinking. Think-aloud protocols, where students verbalize their reasoning while solving problems, have been widely used in research settings~\cite{fonteyn1993description, montague1993middle, lyons2014using}. However, these methods require intensive teacher involvement and can interfere with students' natural problem-solving flow, making them impractical for routine classroom assessment~\cite{henjes2007use}. Written explanations offer a more scalable alternative, yet the quality of students' explanations is highly sensitive to the design of prompts~\cite{chi1989self, renkl2014toward, bisra2018inducing}. To address these limitations, researchers have explored structured approaches, including specialized problem tasks that inherently require students to externalize specific aspects of their reasoning and analytical frameworks to guide teachers' evaluation of student work~\cite{kelly2016analysing}. For example, problem-specific scoring rubrics and carefully designed sub-questions have been proposed to capture students' reasoning at each stage~\cite{schoenfeld2007assessing, jonassen2013assessing, docktor2016assessing, nur2022learning}.

Yet developing such materials for each problem demands substantial expertise and time, making them difficult to scale in practice~\cite{jonassen2013assessing}. Therefore, we aim to explore a scalable approach leveraging LLMs to generate problem-specific diagnostic materials while maintaining pedagogical quality.

\subsection{AI and LLMs in Mathematical Education}
Intelligent tutoring systems (ITSs) have long supported step-level diagnosis of mathematical problem solving. Cognitive tutors trace each step of a student's solution with a cognitive model and provide feedback when the student's reasoning departs from a valid solution path~\cite{Anderson01041995, koedinger1997big}. Systems have also brought similar forms of immediate feedback and formative assessment into classroom settings~\cite{heffernan2014assistments, 10.1145/3698205.3733949}. However, this level of diagnosis typically depends on problems being modeled in advance. Even authoring tools intended for non-programmers require substantial effort for each problem~\cite{aleven2016example, maclellan2022domain}, while identifying which skills a problem assesses still relies heavily on expert judgment~\cite{10.1145/3657604.3662030}. This has largely limited step-level diagnosis to pre-authored problem banks, rather than the problems teachers already use or create for their classes.

Recent advances in large language models (LLMs) have opened new possibilities for supporting mathematics education at scale, spanning question generation, automated scoring, and personalized feedback.
LLMs have demonstrated strong capabilities in generating mathematical problems and assessment items from natural language specifications~\cite{hendrycks2021measuring, 10.3389/feduc.2024.1427502, ariyarathne2025elementary}, with generated items approaching human-authored quality on psychometric criteria~\cite{https://doi.org/10.1111/ijsa.70021}, and recent work has explored decomposing problems into sub-questions to support deeper reasoning~\cite{shridhar2022automatic, yu2025recall, ashok-kumar-lan-2024-improving, NEURIPS2024_9bae399d}. However, most existing approaches focus on generating standalone problems or factual sub-questions, rather than diagnostic sub-questions aligned with specific cognitive stages of problem-solving.
In the domain of assessment, LLMs have been applied to rubric generation and automated scoring of open-ended responses~\cite{cohn2025cotal, wang2025llmarking, 10.1007/978-3-031-98459-4_15, 10.1007/978-3-031-98420-4_15}. Scoring is most reliable when the model is conditioned on an explicit rubric and exemplar answers~\cite{10.1007/978-3-031-98459-4_15}, yet rubrics are often predefined or designed for evaluating complete solutions, rather than being generated adaptively to diagnose student competencies at each problem-solving stage. Recent benchmarks evaluate whether LLM tutors can locate the step at which a student's solution goes wrong~\cite{macina-etal-2025-mathtutorbench}, but they assess the model's judgment rather than producing diagnostic materials a teacher can use.
Beyond generation and scoring, researchers have explored using LLMs to simulate student responses~\cite{lu2024generative}, enabling evaluation of question quality without large-scale human data collection. This direction is relevant for our work, as simulated student errors can inform the construction of level-specific rubrics.
While prior work has demonstrated LLMs' capabilities across these educational tasks in isolation, our research uniquely combines them to support process-level diagnosis. Rather than generating standalone problems or scoring final answers, we automatically produce integrated diagnostic materials---stage-aligned sub-questions and fine-grained rubrics---that cover the full pipeline from material generation to student diagnosis.

\section{Formative Study: Challenges in Diagnosing Problem-Solving Processes in Practice}
To better understand the practical challenges teachers face when diagnosing students' problem-solving skills, we conducted a formative study to examine whether stage-based questions are sufficient for assessing students' reasoning at each problem-solving stage.
Such questions are a natural starting point for scalable diagnosis, as they can be applied across different problems without requiring additional preparation. 
\subsection{Study Design}
We worked with four mathematics teachers who teach students in grades 3 through 6 (3 male, 1 female; mean age = 43.5 years; mean teaching experience = 13.25 years). Drawing on the prior work that decomposes the mathematical problem-solving process into structured stages~\cite{jin2024using}, we used four guiding questions---similar to those in current diagnostic practice:
\begin{enumerate}
    \item \textbf{Parse---Identifying key information}: Underline the important information in the problem.
    \item \textbf{Extract---Planning a solution strategy}: Using the information, describe your plan for solving the problem.
    \item \textbf{Retrieve---Recalling relevant knowledge}: Write down the formulas or concepts needed to solve this problem.
    \item \textbf{Combine---Executing the solution}: Show your complete solution process and arrive at the final answer. 
\end{enumerate}
The teachers then selected grade-appropriate math word problems and presented them to students in a structured worksheet format that incorporated these four questions.

A total of 48 students (distributed across grades 3-6) worked through the three math word problems each, responding to all four questions for each problem. The worksheet included separate response spaces for each stage, allowing students to articulate their reasoning step-by-step. After students completed their work, teachers evaluated each student's performance at each stage of problem solving based on their written responses. Detailed information on participant demographics and a sample worksheet are provided in Appendix~\ref{appendix:formative_study}.

\subsection{Findings from Teacher Interviews}
Following the assessment process, we conducted semi-structured interviews with teachers to understand their experiences diagnosing students' problem-solving skills at each stage, based on their written responses. Teachers consistently reported two major challenges that limited the diagnostic value of this approach.

\textit{\textbf{Limited Diagnostic Value of Student Responses.}}
Teachers reported that generic questions like ``Describe your solution strategy'' frequently failed to surface students' thought processes. One teacher said, ``The questions were the same regardless of the problem, so I couldn't see how students were actually working through it or where their thinking was breaking down.'' Because the questions were not tailored to each specific problem, students' responses tended to be vague and disconnected from their actual problem-solving process. Teachers noted that this issue was particularly problematic in the planning and knowledge-recall stages, where students struggled to externalize their reasoning, indicating that generic questions were insufficient to elicit students' strategy formation and concept application reasoning.

\textit{\textbf{Difficulty in Constructing Stage-Specific Rubrics.}}
Teachers found it difficult to develop clear, consistent evaluation criteria for each problem-solving stage. Because each problem requires different reasoning at each stage, teachers struggled to define what constituted adequate or exemplary responses. One teacher noted, ``I knew what kind of thinking I wanted to see, but it was hard to articulate clear standards that would apply fairly across different students' responses.'' This difficulty was compounded when evaluating multiple problems, as teachers had to mentally adjust their expectations for each problem's context while trying to maintain consistency in their judgments. 

% This lack of specificity in turn made it difficult for teachers to develop clear, consistent evaluation criteria. Because the generic questions were not tailored to specific math word problem contexts, teachers struggled to define what constituted adequate or exemplary responses for each stage. Moreover, they found it difficult to create rubrics that could encompass the diverse range of student approaches and reasoning patterns that emerged in practice. One teacher noted, ``I knew what kind of thinking I wanted to see, but it was hard to articulate clear standards that would apply fairly across different students' responses.'' This difficulty was compounded when evaluating multiple problems, as teachers had to mentally adjust their expectations for each problem's context while trying to maintain consistency in their judgments. 

\subsection{Implications}
These findings highlight a fundamental tension in process-level assessment: while problem-agnostic questions offer broad applicability across problems, they fail to elicit the context-specific reasoning required to solve a given problem, including the problem's constraints, mathematical concepts, and solution strategies. Teachers emphasized that effectively diagnosing problem-solving skills would require questions and rubrics tailored to each problem's context---materials that make explicit what constitutes appropriate reasoning at each stage for that particular problem. However, they also acknowledged that creating such customized materials for every problem would be time-consuming.
Therefore, we explored how LLMs could generate problem-specific questions and rubrics tailored to each problem-solving stage, which can serve as diagnostic materials that teachers need but find too costly to create manually. 

% In the following sections, we describe how we developed a structured problem-solving diagnostic framework and an LLM-based pipeline to address the challenges identified in this study.
\section{\fname{} Framework Development}
The findings of our formative study show that effective diagnosis of students' problem-solving processes requires diagnostic questions that make students' reasoning explicit within the context of a specific problem. However, supporting such context-aligned, process-level diagnosis in a systematic and scalable manner first requires a clear articulation of what skills should be diagnosed at each stage and what appropriate reasoning looks like for each of those skills.

To address this need, we introduce \fname{} (STage-based Evaluation of Problem-solving skills), a structured problem-solving diagnostic framework to support process-level diagnosis of students' math word problem-solving. This framework provides a scaffold for generating context-aligned diagnostic materials, such as stage-specific sub-questions and rubrics, that can be instantiated for individual problems. 
\begin{figure*}[ht]
    \centering
    \includegraphics[width=1\linewidth]{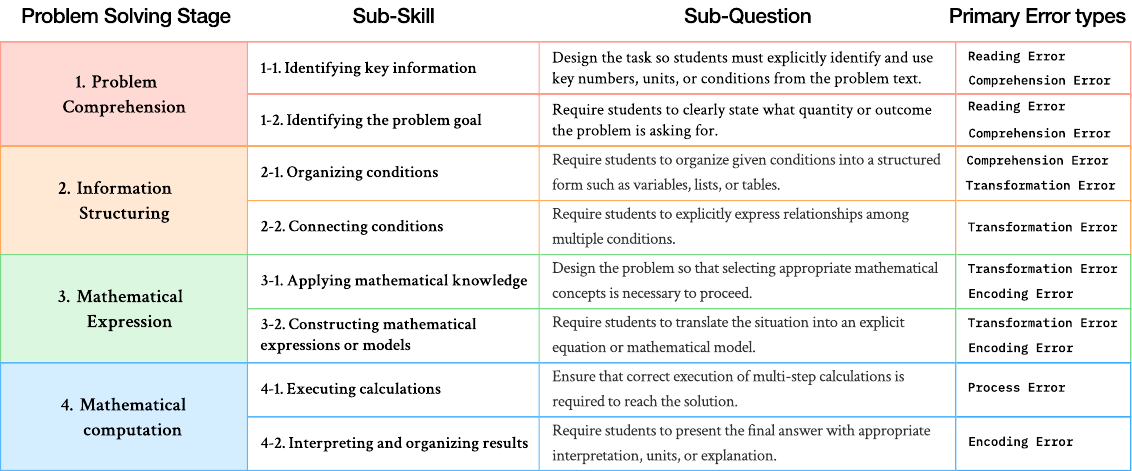}
    \caption{The \fname{} framework for diagnosing math word problem-solving skills. For each problem solving stage, there are two sub-skills, which lead to a sub-question. For each sub-skill, \fname{} also maps primary error types to it, leading to a total of eight sub-questions.}
    \Description{A table-like diagram of the STEP framework showing four problem-solving stages, each with two sub-skills, corresponding sub-question design guidelines, and primary error types. Stage 1 (Problem Comprehension) includes sub-skills 1-1 Identifying Key Information and 1-2 Identifying the Problem Goal, associated with Reading and Comprehension errors. Stage 2 (Information Structuring) includes 2-1 Organizing Conditions and 2-2 Connecting Conditions, associated with Comprehension and Transformation errors. Stage 3 (Mathematical Expression) includes 3-1 Applying Mathematical Knowledge and 3-2 Constructing Mathematical Expressions or Models, associated with Transformation errors. Stage 4 (Mathematical Computation) includes 4-1 Executing Calculations and 4-2 Interpreting and Organizing Results, associated with Process and Encoding errors. Each sub-skill has a corresponding sub-question design guideline describing what the task should require students to do, such as explicitly identifying key numbers and units, stating the problem goal, organizing conditions, expressing relationships, selecting mathematical concepts, constructing equations, executing calculations, and presenting final answers with appropriate interpretation and units.}
    \label{fig:framework}
\end{figure*}

\subsection{Theoretical Foundations}
The development of \fname{} was informed by established problem-solving frameworks in mathematics education, including Polya's four-stage model~\cite{polya2004solve}, Schoenfeld's metacognitive framework~\cite{schoenfeld1983beyond, schoenfeld2014mathematical}, Newman's error analysis~\cite{newman1977analysis}, and mathematics curriculum standards~\cite{ferrini2000principles}. While these frameworks provided conceptual grounding for identifying problem-solving stages, they were developed primarily for instructional purposes rather than diagnostic assessment. Our goal was to operationalize these stages diagnostically, focusing on what evidence teachers should look for and what questions can reliably elicit that evidence. This guided our subsequent co-design process, in which theoretical stages were refined into diagnosable skills.

\subsection{Iterative Framework Development Through Expert Interviews}
Building on the four-stage structure used in the formative study and grounded in the theoretical frameworks described above, we refined the framework through multiple rounds of expert consultation.
We worked with six domain experts, one mathematics education researcher with a Ph.D. in mathematics education, and five mathematics teachers with over 10 years of teaching experience (3 male, 3 female; mean age = 42.3 years, SD = 7.67; mean teaching experience = 16.2 years). 

\paragraph{\textbf{Interview Process.}}
Through individual semi-structured interviews over multiple rounds, we collaboratively developed a more refined framework for process-level diagnosis.
For each session, we prepared a shared document (Google Slides\footnote{https://docs.google.com/presentation/}) containing our current framework draft and example sub-questions generated by the LLM based on that framework. We asked experts to review the materials and leave comments directly in that document. We then conducted individual interviews to discuss their feedback in depth. We specifically discussed (1) whether the current framework captures skills that can be meaningfully diagnosed at each problem-solving stage, (2) what common errors or misconceptions students typically exhibit at each stage, and (3) what characteristics make questions effective at eliciting diagnostic evidence for those skills. Each interview was conducted via Zoom and lasted approximately one hour.

\paragraph{\textbf{Iterative Refinement.}} We began with the mathematics education researcher, whose feedback laid the foundation for the framework structure. The researcher emphasized the importance of distinguishing between general cognitive skills and problem-solving-specific skills, guiding the decomposition of each stage into diagnostically distinct sub-skills---ensuring that each targets a specific reasoning skill that can be independently assessed. We then conducted sessions with the five mathematics teachers, whose classroom experience helped us refine the framework as a practical diagnostic tool and improve the LLM pipeline for generating context-aligned sub-questions. They emphasized that sub-questions should elicit observable evidence of students' thinking at each stage---for instance, rather than simply asking students to perform a calculation, questions should prompt students to articulate how they set up the problem or why they chose a particular approach. This helped clarify what counts as diagnostic evidence at each stage, and informed iterative refinement to the framework and the sub-question generation pipeline. Across these five teacher sessions, we iteratively refined the framework by clarifying stage boundaries, decomposing stages into diagnostically assessable sub-skills, and adjusting prompt-generation guidelines to better elicit students’ intermediate reasoning.
The study was approved by our university’s Institutional Review Board (IRB). The mathematics education researcher received KRW 100,000 in compensation, reflecting their role and involvement in the early stages of framework construction, while each mathematics teacher received KRW 50,000.

% Experts drew on both established problem-solving theories and their practical experience to ensure the framework remained theoretically grounded while addressing practical diagnostic needs.
% This iterative process allowed us to translate broad problem-solving stages into actionable diagnostic specifications—making explicit what reasoning to look for, what errors to watch for, and how to design questions that make students' often-implicit thinking visible.
% This process ensured the framework remained theoretically sound while addressing teachers' practical diagnostic needs—translating broad problem-solving stages into specific guidelines for what reasoning to assess, what errors to watch for, and how to design questions that make students' implicit thinking visible.
% \input{tables/framework}

\subsection{The Final \fname{} Framework}
The resulting \fname{} framework consists of four problem-solving stages and eight sub-skills (\autoref{fig:framework}). Rather than enumerating procedural steps, the framework is designed to support diagnosis by clarifying what reasoning should be elicited and what constitutes diagnostic evidence at different points in the problem-solving process.
% Although the framework presents problem-solving as a sequence of stages, it is not intended to imply a strictly linear process. Experts emphasized that students often revisit earlier stages when encountering difficulties, and that diagnostic evidence may emerge non-linearly across responses. 
Although the problem-solving process is presented as a sequence of stages, each stage and sub-skill is designed to be as diagnostically independent as possible: each can be assessed based on the evidence elicited by its corresponding question, without presupposing successful performance in earlier stages. This design allows teachers to pinpoint where breakdowns occur.
A defining characteristic of \fname{} is its emphasis on early stages of problem solving. Aligning with prior work~\cite{singer2017reading, jin2024using}, experts noted that today's students often struggle more with interpreting problem text and structuring information than with calculations. Reflecting this reality, the framework assigns equal diagnostic importance to early-stage skills.
Together, these elements provide a scaffold for generating context-aligned diagnostic prompts and rubrics, which we operationalize through the LLM-based pipeline described next.
\section{LLM-based Pipeline for Diagnostic Material Generation}
\begin{figure*}
    \centering
    \includegraphics[width=1\linewidth]{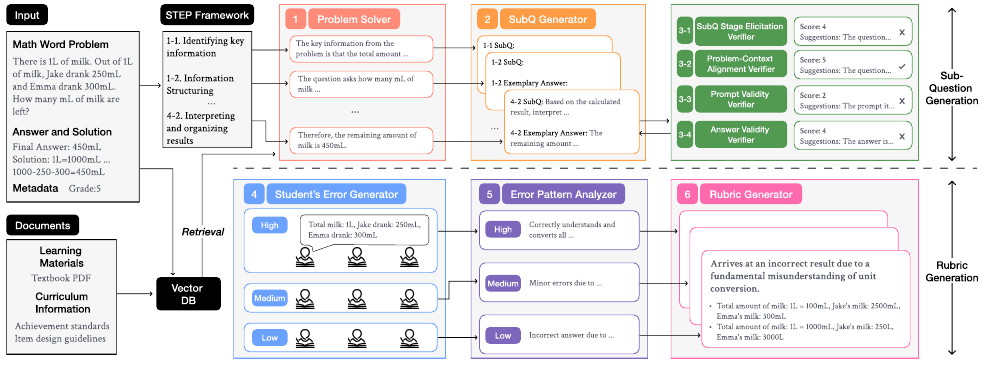}
    \caption{The LLM-based Pipeline for \fname{}. The pipeline transforms a math word problem into verified sub-questions targeting the eight sub-skills, and derives three-tier rubrics from simulated student responses and their error patterns.}
    \Description{A detailed diagram of the LLM-based pipeline for diagnostic material generation, divided into two main sections: Sub-Question Generation (upper) and Rubric Generation (lower). A math word problem, along with its answer, solution, and metadata such as grade level, serves as input to Module 1 (Problem Solver), which produces a step-by-step solution. Separately, teacher-provided documents—learning materials and curriculum information such as achievement standards and item design guidelines—are stored in a vector database and retrieved to ground the generation process in the relevant curriculum content. Module 2 (SubQ Generator) then creates sub-questions aligned to each problem-solving stage, such as SubQ 1-1 asking students to identify key quantities. Four verifiers (3-1 Stage Elicitation, 3-2 Problem-Context Alignment, 3-3 Prompt Validity, 3-4 Answer Validity) evaluate each sub-question and trigger regeneration up to three times if criteria are not met, producing a finalized sub-question with a prompt and exemplary answer. In Rubric Generation, Module 4 (Student's Error Generator) produces simulated student responses at three skill levels (High, Mid, Low) for each sub-question. Module 5 (Error Pattern Analyzer) identifies distinct error patterns across levels. Module 6 (Rubric Generator) constructs a three-tier rubric with scoring criteria and representative student answer examples for each level. The High tier shows criteria for fully correct answers, while Mid and Low tiers include specific error patterns and example incorrect responses.}
    \label{fig:pipeline}
\end{figure*}

We present an LLM-based pipeline (\autoref{fig:pipeline}) that generates diagnostic materials for math word problems, consisting of (1) \textbf{stage-specific sub-questions} that elicit students’ reasoning at each problem-solving stage, and (2) \textbf{fine-grained rubrics} that support structured evaluation of student responses.
This pipeline was developed through iterative collaboration with mathematics teachers, incorporating their pedagogical insights.

\subsection{Sub-Question Generation}
This stage generates a set of diagnostic sub-questions that scaffold students through the four problem-solving stages of the \fname{} framework. The generation process consists of three modules: (1) \textbf{Problem Solver}, which performs CoT-based problem solving, (2) \textbf{Sub-Questions Generator}, which generates sub-questions based on the problem-solving process, and (3) \textbf{Verification and Refinement}, which iteratively refines the sub-questions through internal verification.

This process begins by taking a math word problem, its final answer, and the target grade level as input. When available, a model solution or reference images can be optionally provided. 

\textit{\textbf{Problem Solver. (Figure~\ref{fig:pipeline}-\redbox{1})}}  
Using this information, the pipeline first reasons through the problem step by step, following a chain-of-thought~\cite{wei2022chain} process guided by the \fname{} framework as an interpretive scaffold to identify the types of reasoning required at each stage. The model also identifies the mathematical content domain of the problem \textit{(e.g., Numbers and Operations, Patterns and Relationships, Geometry and Measurement, or Data and Possibility) \cite{MOE2022Math}}. This domain classification enables the pipeline to reference domain- and grade-specific curriculum constraints that inform the design of diagnostic sub-questions. For example, if a problem is categorized within the \textit{Patterns and Relationships} domain for grades 3–4, the curriculum specifies that the problem should not involve complex expressions such as mixed-operation formulas or general-term representations when expressing patterns. In our current implementation, we operationalize these curriculum constraints using Korea’s National Curriculum~\cite{MOE2022Math}. This choice reflects the scope of our study—our dataset consists of Korean math word problems, and our classroom study involves Korean elementary students. However, national curricula are formally specified across educational systems worldwide~\cite{borji2023intended}. Therefore, the pipeline can be adapted to other countries by substituting the relevant curriculum standards.

\textit{\textbf{Sub-Questions Generator. (Figure~\ref{fig:pipeline}-\orangebox{2})}}
Based on the problem-solving process and curriculum constraints, the pipeline transforms the identified reasoning steps into a set of stage-specific sub-questions aligned with the eight \fname{} sub-skills. Sub-questions are generated under design constraints specified in the national curriculum to preserve their diagnostic intent. Each sub-question targets a single \fname{} sub-skill without combining multiple stages, and avoids directive question prompts that specify calculations, procedures, formulas, or expected answer forms.

\textit{\textbf{Verification and Refinement.}}
To improve the educational soundness of generated sub-questions, the pipeline applies a multi-verifier check \cite{Worden1015Scaling}. Four verifiers assess complementary aspects of quality: \textit{stage elicitation (Figure~\ref{fig:pipeline}-\greenbox{3-1}), context alignment (Figure~\ref{fig:pipeline}-\greenbox{3-2}), prompt validity (Figure~\ref{fig:pipeline}-\greenbox{3-3}), answer validity (Figure~\ref{fig:pipeline}-\greenbox{3-4})}. These verifiers operationalize the design constraints described above by checking whether each sub-question satisfies them in practice.
The \textit{stage-elicitation verifier} assesses whether a sub-question targets the intended \fname{} sub-skill. The \textit{context-alignment verifier} evaluates whether a sub-question is grounded in the target problem's context. The \textit{prompt-validity verifier} examines whether a sub-question is appropriate for the target grade level and curriculum scope. The \textit{answer-validity verifier} checks whether the exemplary answer to each sub-question is mathematically correct.
Each verifier scores a sub-question on a 0--5 scale. If any score falls below the maximum-score threshold (5/5), the pipeline regenerates the sub-question using verifier feedback, for up to three iterations. We selected this strict threshold through iterative testing, as lower thresholds allowed wording or content issues to pass verification. Subsequent sub-questions are generated only after the preceding generate--verify--regenerate cycle is completed.
% Each verifier evaluates the quality of a sub-question on a 0–5 scale. If any verifier assigns a score below a predefined threshold, the corresponding sub-question is regenerated. We set this threshold to the maximum score (5/5), based on iterative testing during pipeline development: lower thresholds allowed sub-questions with wording or content issues to pass verification, whereas requiring the maximum score reduced the incidence of such issues.
% The pipeline then regenerates sub-questions conditioned on these suggestions, enabling iterative refinement for up to three iterations until the set of sub-questions satisfies all verification criteria. Each sub-question undergoes a \textbf{generate--verify--regenerate cycle}, and subsequent sub-questions are generated after the previous cycle is completed.

\subsection{Rubric Generation}
This stage generates fine-grained rubrics for evaluating student responses to each sub-question. Rather than relying on generic grading criteria, our pipeline grounds rubric design in diverse student errors across performance levels. 

\textit{\textbf{Students' Error Generator. (Figure~\ref{fig:pipeline}-\bluebox{4})}}
For each of the eight sub-questions, our pipeline generates student responses across three performance levels---low, medium, and high. For both the low and medium levels, three plausible incorrect responses are generated, and for the high level, three correct-answer variations are generated. For each sub-question, the prompt incorporates the correct responses from prior steps, thereby preventing cascading errors from propagating to later stages. The prompts also incorporate skill-specific error characteristics grounded in educational research (e.g., Newman's error types~\cite{newman1977analysis}), such that low-level errors reflect fundamental misunderstandings while medium-level errors reflect partial comprehension with minor gaps. Additionally, grade-level achievement standards and knowledge scope data are incorporated to ensure that generated responses remain grade-appropriate.

\textit{\textbf{Error Pattern Analyzer. (Figure~\ref{fig:pipeline}-\purplebox{5})}}
All nine simulated responses (three per level) for a given sub-question are analyzed to identify qualitative differences across levels. This analysis extracts distinct, non-overlapping error patterns and establishes clear boundaries between low, medium, and high performance. 

\textit{\textbf{Rubric Generator. (Figure~\ref{fig:pipeline}-\pinkbox{6})}}
Using the error pattern analysis, the pipeline constructs a three-tier grading rubric---High (correct), Mid (partial understanding), and Low (fundamental error)---with criteria and representative examples drawn directly from the simulated responses. This approach ensures that rubric criteria are specific and actionable, being grounded in concrete error patterns observed by the Error Pattern Analyzer. 

All modules in the pipeline employ OpenAI's \texttt{GPT-4o} model to leverage its multimodal capabilities for processing problems that may include diagrams or visual elements. Detailed prompts for each module are provided in the supplementary materials.
\section{Two-stage Evaluation}
To examine how our \fname{} framework and LLM pipeline function in practice, we conducted a two-stage evaluation. Stage 1 involved an \textbf{expert review} to gather feedback on the mathematical and pedagogical validity of the diagnostic materials generated by the LLM pipeline across diverse problems. Based on the expert feedback, we refined the pipeline and developed \sysname{}, an end-to-end teacher-in-the-loop system that supports the workflow from diagnostic material generation to student diagnosis. Stage 2 then involved a month-long classroom deployment of \sysname{} with \numTeachers{} teachers and \numStudents{} students, examining how the \fname{} framework and \sysname{} system function in real educational contexts to support process-level diagnosis of students’ problem-solving.

\subsection{Stage 1: Expert Review}
We conducted an expert review with mathematics teachers to gather structured feedback on diagnostic materials generated by the LLM pipeline based on \fname{}. The review aimed to (1) understand how teachers interpret the pedagogical appropriateness of the generated sub-questions and rubrics, and (2) identify practical considerations and refinements for classroom use.

\subsubsection{Review Protocol}

\textit{\textbf{Dataset.}}
We used math problem data from the AI Hub~\cite{aihub}, which provides curriculum-aligned problems with problem statements, correct answers, model solutions, and difficulty labels across grade levels. We filtered for medium-to-high difficulty math word problems at grades 3--6 that require multi-step reasoning, and sampled 36 problems covering diverse mathematical content domains. Due to the smaller pool of eligible lower-grade problems, we combined grades 3 and 4 and sampled 12 problems for this band, and 12 problems each for grades 5 and 6. Details on the sampled problems are in Appendix (\autoref{tab:tech_eval_problems}).

\textit{\textbf{Participants.}} 
Six mathematics teachers with at least 5 years of teaching experience were recruited through open calls and snowball sampling (3 male, 3 female; mean age = 38.2, mean teaching experience = 11.93 years), with two assigned to each grade band (grades 3--4, 5, and 6). Each teacher evaluated 12 problems, each consisting of eight stages, resulting in 96 sub-questions and their accompanying rubrics. Details are in Appendix (\autoref{tab:expert_review_participants}).

\textit{\textbf{Procedure.}} The review consisted of three phases: (1) a 30-minute orientation introducing the research purpose, \fname{} framework, and evaluation criteria, (2) an asynchronous evaluation session of approximately 2.5 hours in which teachers assessed the generated diagnostic materials using given metrics, and (3) a one-hour follow-up interview to discuss evaluation reasoning and suggest improvements for practical classroom use. Each participant was compensated 160,000 KRW for their time.

\textit{\textbf{Review Dimensions.}}
Drawing on prior work for assessing the validity and usability of educational assessment materials~\cite{brookhart2013create,reddy2010review}, we defined structured review dimensions to guide teacher feedback. Teachers reviewed the generated materials at two levels of granularity: each of the eight sub-skills individually, and the full set generated for a given problem.
At the sub-skill level, teachers reviewed the materials generated for each of the eight \fname{} sub-skills on three dimensions: whether the sub-question and rubric target the intended sub-skill for diagnosis (Construct Validity), whether their complexity, information load, and phrasing suit the target grade level (Level Appropriateness), and whether the exemplary answer and rubric criteria are mathematically correct (Mathematical Accuracy).
At the problem level, teachers reviewed the full set of materials generated for a given problem, examining how the \fname{} framework was concretely instantiated in a specific problem-solving context. They judged whether the materials naturally connect to the problem context and guide students through the solving process coherently (Problem-Context Alignment), and whether they sufficiently cover the full problem-solving process required by the problem (Comprehensiveness).
For all dimensions, teachers made binary (Yes/No) judgments alongside free-form comments, encouraging clear decisions while capturing the reasoning behind them and surfacing disagreements for follow-up. The exact review questions are in Appendix~\ref{appendix:review_questions}.

\textit{\textbf{Supplementary Review.}}
For each grade band, two teachers evaluated the same set of problems. When they disagreed, two additional mathematics teachers who had not participated in the primary review served as third-party tiebreakers, each assigned half of the conflicting cases, to provide supplementary pedagogical perspectives and reach a final determination.

\subsubsection{Expert Review Result} 
Overall, the expert review suggested that materials generated by the LLM pipeline, which operationalizes the \fname{} framework, were generally viewed as pedagogically valid and structurally coherent, while also revealing specific areas for refinement. Below, we discuss the results for each component in detail. Quantitative results are shown in ~\autoref{tab:expert_review_results}.

\begin{table}[t]
\centering
\begin{tabular}{llc}
\toprule
\textbf{Component} & \textbf{Metric} & \textbf{Score} \\
\midrule
\multirow{3}{*}{\shortstack[l]{Sub-skill Level}} 
 & Construct Validity& 91.32\% \\
 & Level Appropriateness & 80.21\% \\
 & Mathematical Accuracy & 90.97\% \\
\midrule
\multirow{2}{*}{\shortstack[l]{Problem Level}} 
 & Problem-Context Alignment & 72.22\% \\
 & Comprehensiveness & 77.78\% \\
\bottomrule
\end{tabular}
\caption{Expert review results. Score indicates the proportion of Yes judgments.}
\Description{A table of expert review results, where score indicates the proportion of Yes judgments. At the Sub-skill Level: Construct Validity 91.32 percent, Level Appropriateness 80.21 percent, Mathematical Accuracy 90.97 percent. At the Problem Level: Problem-Context Alignment 72.22 percent, Comprehensiveness 77.78 percent.}
\label{tab:expert_review_results}
\end{table}
\textit{\textbf{Sub-skill Level Result.}}
Teachers rated the generated materials positively across all three dimensions: Construct Validity (91.32\%), Mathematical Accuracy (90.97\%), and Level Appropriateness (80.21\%). Qualitatively, teachers affirmed that the stage-based structure of \fname{} was well-reflected in the generated materials, and that the decomposition into sub-skills made diagnostic intentions explicit. The relatively lower rating for Level Appropriateness reflected occasional phrasing issues: three teachers (E2--E3, E6) suggested that some prompts could be simplified for elementary students, and six (E2, E4--E8) noted wording that could be better aligned with grade-level curriculum. In particular, E6 and E7 proposed uploading the learning materials they already use in class, so that the generated materials would draw on the terms and solution methods their students have encountered. E5 also remarked that correcting such expressions directly would be simpler than requesting a regenerated version from the AI. Teachers also found the rubrics' inline student-response examples useful for anticipating how students reason at each stage, though several noted that their specificity occasionally drew attention to particular cases, making score boundaries less intuitive (E2--E4, E6--E8).

\textit{\textbf{Problem Level Result.}}
For the full set of materials generated for each problem, 72.22\% were rated as coherently connected to the problem context (Problem-Context Alignment), and 77.78\% as comprehensively covering the problem-solving process (Comprehensiveness). Five teachers (E1--E3, E5--E6) agreed that the sets guided students through the problem-solving stages in a coherent and educationally meaningful way, supporting the validity of \fname{} across diverse problems. At the same time, four teachers (E1--E2, E5--E6) noted that a single phrasing or wording misalignment could lower the rating for an entire set. This yielded slightly lower scores than at the sub-skill level, pointing to individual sub-question quality as an area for refinement.
\begin{figure*}
    \centering
\includegraphics[width=1\linewidth]{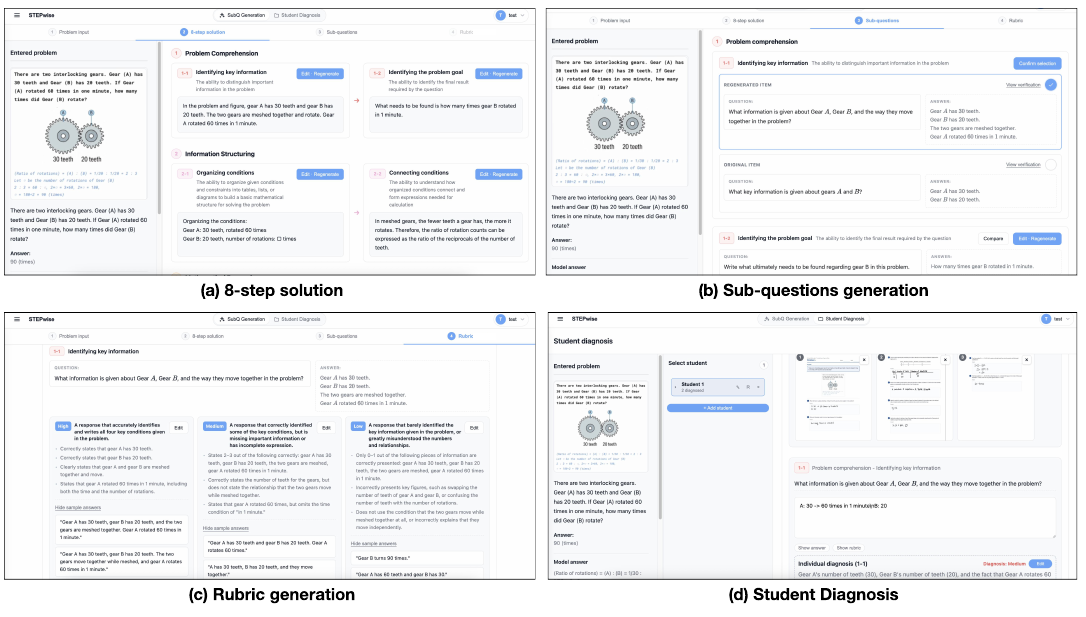}
    \caption{The \sysname{} system interface. (a) The 8-step solution view shows the input math word problem alongside the pipeline's step-by-step solving process, organized under the STEP framework's four stages. (b)The sub-question generation view shows both the original LLM-generated sub-question and a regenerated version for each sub-skill. Teachers can edit questions, view verification results, provide feedback, or hide questions. (c) The rubric generation view for the same sub-skill displays a three-tier rubric (High, Medium, Low) with scoring criteria and student example responses. Teachers can edit criteria and regenerate rubrics. (d) The student diagnosis view lets teachers upload a student's submitted responses, which the LLM then diagnoses against the previously generated rubrics.}
    \Description{Four screenshots of the STEPwise system, arranged in a 2x2 grid and labeled (a) through (d). (a) 8-step solution: the entered math word problem about two interlocking gears is shown on the left, with the pipeline's step-by-step solution organized into four stages on the right (Problem Comprehension, Information Structuring, Mathematical Expression, Mathematical Computation), each stage showing one or two sub-skills (e.g., Identifying key information, Identifying the problem goal, Organizing conditions, Connecting conditions) with their generated content and Edit/Regenerate buttons. (b) Sub-question generation: the same gear problem is shown alongside a sub-question generation panel for the Problem Comprehension stage, displaying an original sub-question and a regenerated sub-question for the "Identifying key information" sub-skill, each with a question, answer, and a "View verification" link, plus a "Confirm selection" button; a second sub-skill, "Identifying the problem goal," is shown below with Compare and Edit/Regenerate options. (c) Rubric generation: for the "Identifying key information" sub-skill, three rubric tiers are shown side by side — High, Medium, and Low — each with a description of the scoring criteria, a bulleted list of specific conditions, and a "Show sample answers" link with example student responses. (d) Student Diagnosis: the entered problem is shown on the left, a student selection panel is shown in the middle where a student's uploaded response images are shown, and below, an individual diagnosis panel for the "Identifying key information" sub-skill shows the student's submitted answer, an LLM-generated diagnosis label based on the rubric (e.g., "Medium"), and an Edit button.}
    \label{fig:system}
\end{figure*}

\subsection{Stage 2: Classroom Study}
\subsubsection{\sysname{} System Development}
The expert review informed two directions for making the pipeline work in real classrooms. First, teachers suggested that the generated sub-questions and rubrics should better reflect the actual curriculum. In response, we added a retrieval-augmented generation (RAG) ~\cite{lewis2020retrieval} component through which teachers upload their own textbooks or learning materials. Given an input problem, the pipeline retrieves the most relevant passages from the uploaded materials via embedding-based similarity search, and incorporates them into its reasoning process when generating sub-questions and rubrics. This ensures that the resulting terminology, solution methods, and approaches align with the curriculum content actually covered in that grade and classroom. Second, teachers noted that adjusting phrasing themselves would be simpler than requesting a regenerated version from the AI (E5). We therefore adopted a teacher-in-the-loop design~\cite{rodriguez2018teacher}, in which teachers can directly review, edit, and provide feedback on generated materials before deploying them to students.

The resulting system, \sysname{}, supports the full workflow of diagnosing students' problem-solving skills, from diagnostic material generation to student response assessment. Teachers begin by inputting a math word problem, optionally along with the textbooks or learning materials used in their class, upon which the system first solves the problem step-by-step following the \fname{} framework (Figure~\ref{fig:system}a), and then generates stage-aligned sub-questions (Figure~\ref{fig:system}b) and corresponding rubrics (Figure~\ref{fig:system}c). At each stage of the generation process, teachers can review and directly edit the outputs, provide natural language feedback to regenerate specific parts, or request full regeneration---preventing errors from propagating to subsequent stages. Alternatively, once teachers feel that the generation has stabilized beyond a certain point, they can opt to automatically generate the remaining stages in a single batch. Teachers can then deploy the finalized materials to students, for example by exporting them as a PDF worksheet. Once students submit their responses, \sysname{} supports teachers in assessing each response against the rubrics (Figure~\ref{fig:system}d) and generates a diagnostic report summarizing students' problem-solving processes at each stage (see Figure~\ref{fig:report-sample} in the Appendix for a sample report).

\subsubsection{Classroom Study Design and Procedure}
\textit{\textbf{Participants.}}
We recruited \numTeachers{} mathematics teachers (CT1--CT5) through an open call (all female; mean age = 41.2, SD = 7.7), together with \numStudents{} of their students spanning grades 3 through 9. While the pipeline had been reviewed for grades 3--6, several teachers asked to use \sysname{} with their middle-school classes as well; we allowed this and report their experience as an exploratory extension beyond the reviewed grade range. Informed consent was obtained from the teachers, and assent was obtained from students with parental consent. This study was approved by the IRB of our university.

\textit{\textbf{Procedure.}}
To examine how \sysname{} is used in real educational contexts and what educational value it offers, we deployed the system to teachers and had them use it in their own classes for approximately one month. After a 40-minute orientation covering the research goals, the \fname{} framework, and the \sysname{} workflow, teachers used the system on their own: they selected problems of their choosing, reviewed and edited the generated sub-questions and rubrics, deployed the materials to their students, and submitted the collected responses to \sysname{}, which scored them against the rubrics and produced a diagnostic report for each student. How the materials were used---in class, as homework, or both---was left to each teacher. Teachers ran this cycle problem by problem, so each student's diagnostic report accumulated over the month rather than reflecting a single session. We held a 30-minute check-in interview midway through the deployment, and at the end teachers completed a survey including the System Usability Scale (SUS)~\cite{bangor2008empirical} and open-ended questions, followed by a 30-minute final interview. Students completed a brief survey of four dimensions---usefulness of the step-by-step structure, ease of expressing the reasoning process, self-awareness of their strengths and weaknesses, and willingness to use it for future learning---on a 5-point Likert scale, with one open-ended question comparing this approach to their typical math problem-solving experience (see Appendix~\ref{appendix:student_survey}).
% Students’ written responses were digitized by the research team using OCR and checked for accuracy before being entered into the system.

\subsubsection{Classroom Study Results}
\begin{figure}
    \centering
    \includegraphics[width=\linewidth]{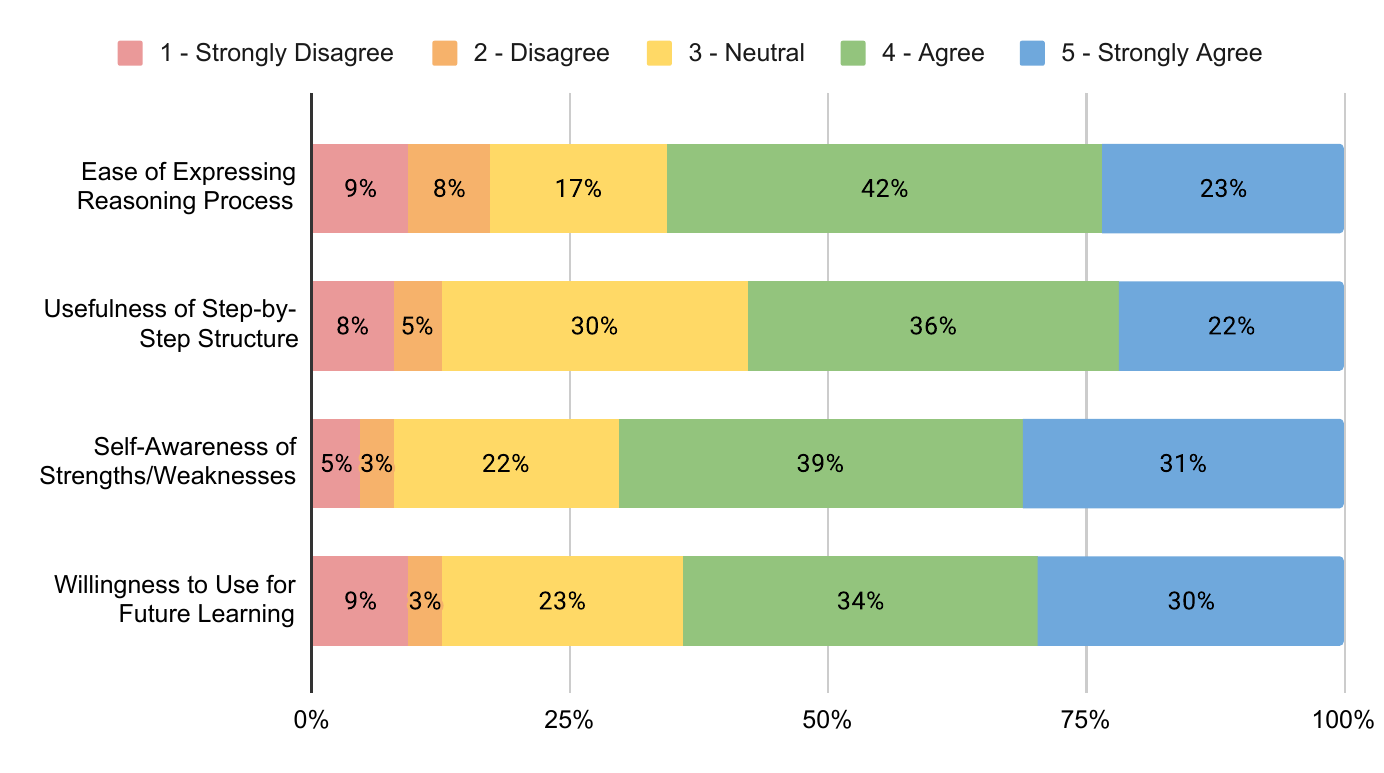}
    \caption{Student survey responses after classroom deployment ($N=64$). Bars show response distributions on a 5-point Likert scale; 58--70\% of students agreed or strongly agreed across the four items.}
    \label{fig:classroom_likert}
    \Description{A horizontal stacked bar chart of student survey results (N=64) across four dimensions, each rated on a 5-point scale from Strongly Disagree to Strongly Agree, ordered top to bottom. Ease of Expressing Reasoning Process: 9 percent strongly disagree, 8 percent disagree, 17 percent neutral, 42 percent agree, 23 percent strongly agree. Usefulness of Step-by-Step Structure: 8 percent strongly disagree, 5 percent disagree, 30 percent neutral, 36 percent agree, 22 percent strongly agree. Self-Awareness of Strengths and Weaknesses: 5 percent strongly disagree, 3 percent disagree, 22 percent neutral, 39 percent agree, 31 percent strongly agree. Willingness to Use for Future Learning: 9 percent strongly disagree, 3 percent disagree, 23 percent neutral, 34 percent agree, 30 percent strongly agree.}
\end{figure}

\paragraph{\textbf{Scalability of Material Generation and Diagnosis.}}
Over the month, the \numTeachers{} teachers prepared diagnostic materials for 90 problems in total, amounting to 720 sub-questions and accompanying rubrics, and reported no notable difficulty in using the system (mean SUS score~\cite{bangor2008empirical} of 83.5). As teachers chose the problems themselves, the materials reflected what each class was actually working on, and varied in topic and difficulty---ranging from regular curriculum problems to competition-level ones. Teachers attributed both the volume and this freedom of choice to the fact that the only input required was a problem they already intended to use: the system produced the stage-specific sub-questions and rubrics for whatever they brought to it, and scored student responses against them once collected. Teachers described manually authoring eight sub-questions and corresponding three-tier rubrics for every problem as impractical alongside their regular teaching. They valued being able to generate diagnostic materials for any problem they were already using in their classes. They therefore viewed \sysname{} as enabling a form of process-level diagnosis that they could not otherwise sustain. CT4 noted that, among the systems she had used, its questions were the closest to how questions are posed in her own classroom.

\paragraph{\textbf{Making the Problem-Solving Process Visible.}}
Teachers consistently described the central benefit of \sysname{} as observing the flow of student thinking rather than the correctness of a final answer. The stage-based decomposition let them check students' comprehension, conceptual knowledge, and logical reasoning separately, and identify how far a student had progressed and where their reasoning broke down. In practice, when students cannot solve a problem, they often simply say ``I don't know,'' and the teacher must probe through follow-up questions to locate the breakdown; \sysname{} made this visible without additional diagnostic questioning. Teachers also valued the three-tier rubrics as an explicit marking standard: having High, Medium, and Low criteria spelled out gave them a clear reference point when giving written feedback on student work.
This visibility changed what teachers knew about their own students. CT5 reported being surprised that a student they had vaguely assumed was doing well turned out to be stuck at a particular stage, and that seeing this made it much clearer what support to provide for that student. Teachers also noted that the student diagnosis reports let them identify which students were struggling at a specific stage, such as problem comprehension. CT2 and CT5 further pointed out that reports of this kind are directly usable in parent consultations, since they let a teacher explain concretely where a student struggled rather than simply saying that the student is doing well or poorly overall.

\paragraph{\textbf{Student Experience.}}
Students rated the approach moderately positively across all four survey dimensions (means between 3.59 and 3.89 out of 5; see \autoref{fig:classroom_likert}), with distributions skewed toward agreement alongside a persistent minority who disagreed. The highest-rated dimension was self-awareness of their strengths and weaknesses ($M=3.89$, $SD=1.04$), suggesting that working through the stages made students' own reasoning gaps visible to them. This was followed by willingness to use the approach for future learning ($M=3.72$, $SD=1.20$), perceived usefulness of the step-by-step structure ($M=3.63$, $SD=1.20$), and ease of expressing their reasoning process ($M=3.59$, $SD=1.12$)---the lowest-rated dimension, consistent with teachers' observation that articulation was the central difficulty.
In open-ended responses, students reported that breaking problems into stages helped them solve more carefully and write out their thinking more explicitly than usual, and that using what they had established in earlier stages to answer later ones helped them see the structure of the problem. At the same time, several reported that writing out each step took more time than usual, since they needed to explain their thinking instead of just giving an answer.

\paragraph{\textbf{How Students Engaged with Stage-Based Sub-Questions.}}
A finding teachers repeatedly described as unexpected was that decomposing a problem into stages did not make it easier for students. Teachers had assumed that presenting a problem step by step would lower its difficulty, but many students found the stage-based sub-questions harder than the original problem. The difficulty lay not in the mathematics but in the act of articulation: breaking one's thinking into steps and expressing it in sentences or expressions was itself the obstacle. Even students who ordinarily solve such problems well became stuck, and students who had never written out explanations sometimes gave up partway. Teachers observed that students across the board tried to jump directly to an answer rather than dividing their thinking into parts.
Grade-level differences were pronounced. CT2 contrasted her elementary students in grades 3--4, who tried to write down what they knew in as much detail as they could even on competition-level problems, with her middle-school students, who more often judged a problem to be difficult from the outset and resisted recording their reasoning one stage at a time. Teachers read this as evidence that differences in disposition and in the ability to express one's thinking were more visible than differences in mathematical ability, and concluded that these are themselves things to be taught rather than prerequisites to be assumed. This asymmetry echoes prior deployments of LLM-based scaffolding in K--12 classrooms, where a uniform step-by-step structure benefited students overall while placing a disproportionate burden on those with lower proficiency~\cite{myung2026scaffolding}.

Even so, teachers observed changes in how students engaged with difficult problems. Although the level of detail in students' written solutions did not change dramatically, students spent longer thinking before writing, and when they encountered a hard problem, they re-read it and considered alternative approaches instead of giving up. CT1--CT4 also noticed a change in how students asked for help: rather than asking for the answer, some students began asking questions such as ``which part should I think about?'' or requesting a small hint, and students developed the habit of first grasping what a question was asking before writing their solution. Teachers therefore characterized the sub-questions as not only a diagnostic instrument but a learning experience in themselves, giving students practice in working through a single problem thoroughly.

\paragraph{\textbf{Adapting the Decomposition to Problems and Students.}}
Teachers regarded stage-based decomposition as pedagogically sound overall: dividing a problem into stages let them see how far a student's understanding extended, and the stage-specific questions themselves gave students a direction for what to consider next. They saw it as particularly valuable for students who struggle with mathematics or with sustained thinking, since it replaces the blanket judgment that a student ``cannot solve problems'' with a specific stage at which difficulty arises. Within this, they pointed to cases where the granularity could flex. CT1 noted that for some problems the intermediate stages already revealed enough about a student's reasoning, so the full eight stages were not always necessary. Teachers also observed that the students who stood to benefit most were sometimes the ones who found eight separate sub-questions most burdensome, and suggested merging adjacent stages into fewer, broader sub-questions for them. For students with strong or creative reasoning, they cautioned that a fixed sequence of stages can channel thinking along the provided path rather than letting students devise their own.
Teachers also saw room for sub-questions that extend thinking rather than lead to the next answer. CT1 observed that some sub-questions read as guiding students toward the following step, and wanted more questions asking why students thought as they did, what alternative solutions exist, or how the situation would change if a condition were altered. A parallel point applied to the rubrics: teachers found the criteria clearly differentiated across score levels, but CT2 felt they were oriented toward stage-wise correctness and wanted criteria that could also credit diverse or creative solution paths. Finally, teachers found the diagnostic reports useful for grasping a student's overall state, while CT2 and CT4 wanted them to be more specific and more actionable---which types of errors recurred and offering guidance for future instruction.

\section{Discussion}
We discuss the broader implications of our findings across three dimensions: students' self-diagnosis, personalized learning, and scaling toward automated deployment. We also reflect on the limitations of the current work and directions for future research.

\subsection{Supporting Student Self-Diagnosis}
Although \sysname{} was originally designed to support teachers, our classroom study suggests its potential to support student self-diagnosis as well. Self-awareness of their own strengths and weaknesses was the highest-rated dimension in the student survey ($M=3.89$), and several students noted that the stage-based structure helped them articulate their thinking more explicitly than usual. Teachers likewise observed that students began re-reading problems, considering alternative approaches, and asking which part to think about rather than asking for the answer. This points to a broader opportunity: because each sub-question targets a diagnostically distinct sub-skill within the \fname{} framework, stage-based sub-questions may naturally scaffold metacognitive reflection \cite{flavell1979metacognition}, helping students recognize precisely where their reasoning breaks down without requiring explicit teacher intervention. This aligns with research on self-regulated learning, which emphasizes the importance of students monitoring and evaluating their own cognitive processes as a foundation for deeper learning~\cite{zimmerman2002becoming}. In such a setting, \sysname{} could present stage-specific sub-questions while withholding the full solution, allowing students to work through each reasoning stage and develop awareness of where their thinking begins to falter. Future work could explore designs that provide students with immediate feedback at each stage, enabling them to self-diagnose their reasoning gaps and take greater ownership of their learning process.

\subsection{From Diagnosis to Personalized Learning}
Our findings open a broader opportunity beyond diagnosis: leveraging stage-level reasoning analysis to support personalized learning. Rather than treating diagnostic results as static assessments, stage-specific performance profiles could inform targeted instructional interventions tailored to each student's reasoning gaps. For instance, a student who possesses strong \textbf{problem comprehension} skills could skip that stage ahead, while a student who struggles with\textbf{ information structuring} could receive additional scaffolded practice on organizing given conditions into variables or tables. Such an adaptive approach would move beyond one-size-fits-all diagnosis toward more personalized support, and connect naturally to the broader literature on mastery-based and adaptive learning systems~\cite{vanlehn2011relative}.

Our classroom study points to a further dimension of personalization: not only which stage to support, but how finely to decompose it. Teachers judged stage-based diagnosis most valuable for students who struggle, yet those students found eight separate sub-questions the most burdensome, and teachers suggested merging adjacent stages for them. Prior classroom deployments report a similar asymmetry, where step-by-step LLM scaffolding improved students' output overall yet demotivated lower-proficiency students and increased their reliance on the system~\cite{myung2026scaffolding}. At the other end, teachers cautioned that a fixed sequence of stages may channel students with strong or creative reasoning along the provided path rather than letting them devise their own. Granularity is therefore not a fixed property of the framework but a parameter to be set per student and per problem: stage-level profiles could inform how many sub-questions a student receives, and which stages are merged, in addition to what content to reteach. Future work could explore how to translate stage-level diagnostic results into concrete personalized instructional interventions, and how much decomposition each student needs to make their reasoning visible without overwhelming them.

\subsection{Toward Scalable and Automated Diagnosis}
\sysname{} currently relies on a teacher-in-the-loop structure in which teachers review and refine generated materials before deployment. This design ensures reliability, and teachers in our classroom study did not find the review process burdensome. Still, scaling this approach to broader contexts, such as large classrooms or online learning platforms, would benefit from further reducing teacher overhead. In our current implementation, \sysname{} already takes a step in this direction: rubrics generated by the pipeline are used directly by an LLM to automatically score student responses, and the resulting scores are aggregated into a diagnostic report that summarizes each student's performance across reasoning stages. This end-to-end flow---from problem input to scored diagnostic report---demonstrates the feasibility of automating the diagnosis process with minimal teacher involvement. 

Further scaling involves two bottlenecks. First, \textit{rubric quality}: teachers in our expert review noted that rubric criteria occasionally narrowed to specific cases, which may reflect limited variability in the responses simulating students' errors. Simply increasing the number of simulations would add computational overhead without guaranteeing broader coverage. A more promising direction is to incorporate real student data as it accumulates through deployment---abstracting recurring error patterns and reasoning strategies across problems and stages---and using these empirically grounded patterns to inform the simulation process itself. This would enable the pipeline to generate richer, more representative rubrics even for new problems, improving diagnostic accuracy without requiring per-problem data collection. Second, \textit{response collection}: in our classroom study, students completed worksheets on paper, which allowed them to express their reasoning naturally. To make this compatible with digital scoring, \sysname{} allowed teachers to upload photos of these paper worksheets for automatic recognition. However, this workaround does not fully resolve the challenge of scaling response collection, as uploading and managing files for a much larger number of students would itself become a burden for teachers. Therefore, future work should explore computer-based input formats that preserve this expressiveness---for example, adaptive block-based interfaces that present relevant mathematical operators without prescribing the sequence of operations.

Automating diagnosis also raises the question of what a diagnostic report should contain. Teachers in our classroom study found the reports useful for grasping a student's overall state, but wanted them to be more specific and more actionable: which types of errors recurred across problems and what concepts or problem types to teach next. They also wanted results to accumulate, so that repeated diagnoses would show how a student's reasoning changes over time rather than reporting each worksheet in isolation. Both point beyond per-worksheet scoring toward a longitudinal record of stage-level performance, which would also supply exactly the empirically grounded error patterns described above.

\subsection{Limitations and Future Work}
This work has several limitations that point to directions for future research. First, stage-based sub-questions do not merely observe the problem-solving process---they structure it. What our approach diagnoses is therefore a student's reasoning under a given decomposition, which may differ from how the same student would approach the problem unaided. Teachers' concern that a fixed stage sequence may channel students with strong reasoning along the provided path is one manifestation of this, and future work should examine how the presence and granularity of sub-questions shape the very process being diagnosed. Second, while our study captured students' immediate experience with the diagnostic materials, we did not measure whether engagement with stage-based sub-questions leads to gains in problem-solving performance, nor did we follow the instructional decisions teachers made based on the accumulated diagnostic reports. Longitudinal studies that trace this full loop---from diagnosis to targeted instruction to subsequent performance---would be an important next step in establishing the educational efficacy of this approach. Third, \sysname{}'s diagnosis step supports the process-level assessment workflow, but we did not treat it as a fully validated diagnostic instrument in this work: we did not systematically compare \sysname{}'s scores against teachers' own scoring of the same responses, an important next step for establishing diagnostic accuracy at scale. Still, teachers raised no concerns about the correctness of the generated reports over the course of the deployment.
Together, these directions would help establish the broader educational value of LLM-generated diagnostic materials as a scalable tool for supporting process-level assessment in mathematics education.
\section{Conclusion}
We investigated how fine-grained diagnosis of students’ mathematical problem-solving processes can be structured and supported in classroom practice. Our formative study identified two practical challenges: generic stage-based questions often failed to elicit problem-specific reasoning, and teachers lacked clear criteria for assessing students’ responses consistently. To address these challenges, we developed \fname{}, which refines four sequential problem-solving stages into eight diagnostic sub-skills, and an LLM-based pipeline that generates problem-specific sub-questions and rubrics. An expert review found strong construct validity and mathematical accuracy while revealing limitations in grade-level and problem-context alignment, motivating the teacher-in-the-loop design of \sysname{}. In a month-long classroom deployment, teachers reported that students’ stage-by-stage responses helped them locate reasoning breakdowns, while students generally valued the support but varied in their ability to articulate each stage. Together, these findings identify both opportunities and design requirements for making fine-grained, process-level diagnosis more practical at classroom scale.

% We introduced an LLM-based approach to diagnosing students' problem-solving skills at each reasoning stage in math word problems. To this end, we first developed \fname{}, a structured diagnostic framework decomposing problem-solving into four stages and eight sub-skills, in collaboration with six mathematics education experts, and built an LLM-based pipeline that automatically generates stage-specific sub-questions and rubrics tailored to individual problems. Our expert evaluation and classroom study demonstrated both the validity of the generated materials and their practical value: teachers could pinpoint where students' reasoning breaks down, and students were able to articulate their thinking more explicitly than usual. We hope this work serves as a step toward scalable, process-level assessment that supports not only diagnosis, but the development of structured mathematical reasoning itself.

\bibliographystyle{ACM-Reference-Format}
\bibliography{references}
\clearpage
\appendix
\raggedbottom
\section{Appendix}
\subsection{Formative Study Details}\label{appendix:formative_study}

\begin{figure*}[p]
\centering
\includegraphics[
  width=0.8\textwidth,
  height=\textheight,
  keepaspectratio
]{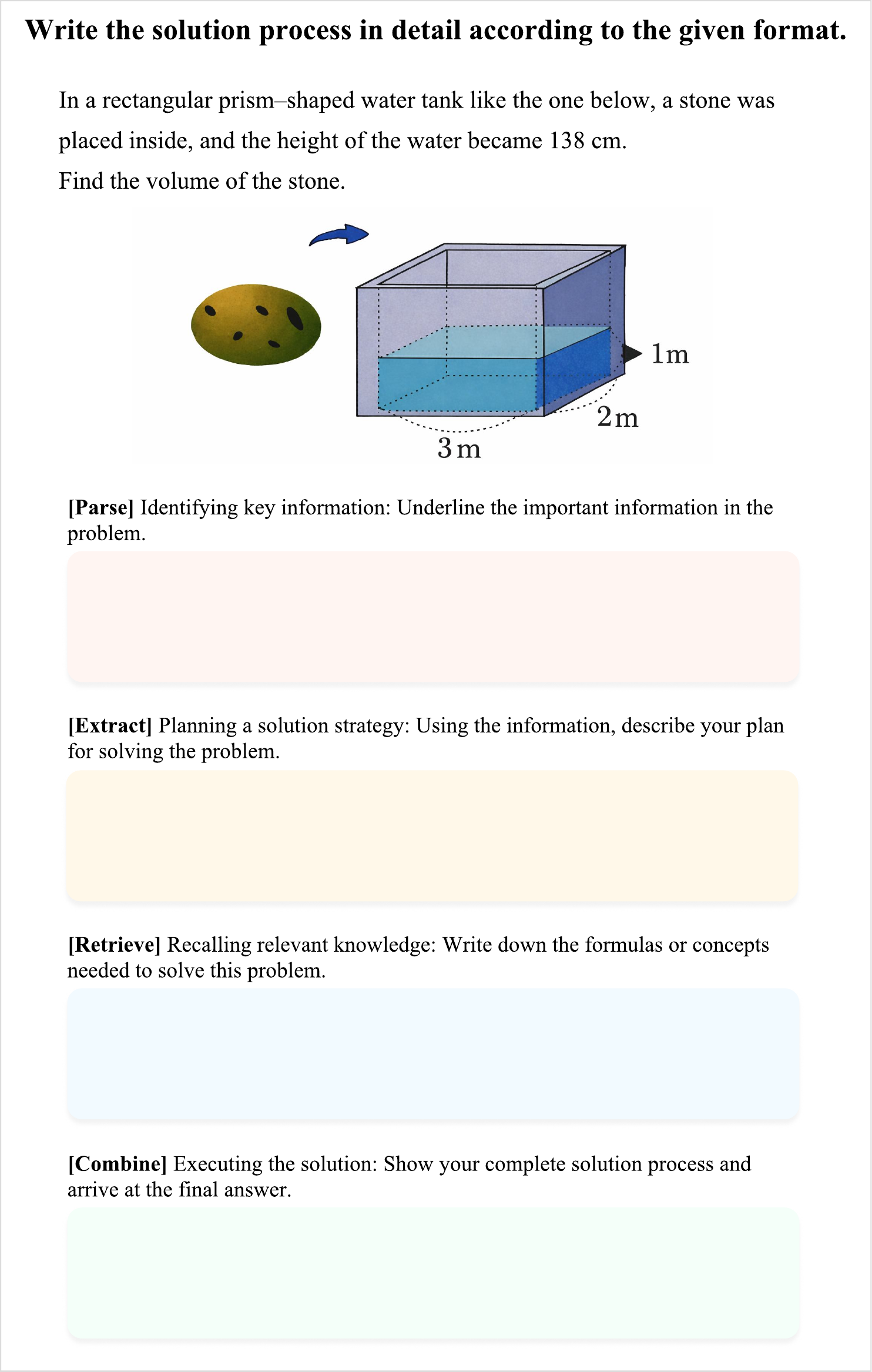}
\caption{A sample worksheet we used for the formative study.}
\label{fig:formative_worksheet}
\Description{A sample worksheet used in the formative study, titled "Write the solution process in detail according to the given format." It presents a word problem about a stone placed in a rectangular prism-shaped water tank measuring 3 meters by 2 meters by 1 meter, where the water height became 138 centimeters, asking students to find the volume of the stone. Below the problem are four labeled response boxes corresponding to the four guiding questions: Parse (Identifying key information), Extract (Planning a solution strategy), Retrieve (Recalling relevant knowledge), and Combine (Executing the solution), each left blank for student responses.}
\end{figure*}

Table~\ref{tab:formative_teacher_demographics} summarizes the demographics of teachers who participated in the formative study. Figure~\ref{fig:formative_worksheet} shows a sample worksheet used during the study to elicit teachers' problem-solving processes and instructional strategies.

\begin{table}[H]
\resizebox{\columnwidth}{!}{%
\centering
\begin{tabular}{lccc}
\toprule
\textbf{Teacher ID} & \textbf{Age} & \textbf{Gender} & \textbf{Teaching Exp. (yrs)} \\
\midrule
Formative-T1 & 45 & Male & 25 \\
Formative-T2 & 44 & Male & 3 (+15 yrs in assessment question design) \\
Formative-T3 & 48 & Male & 20 \\
Formative-T4 & 37 & Female & 5 \\
\bottomrule
\end{tabular}
}
\caption{Teacher demographics for formative study (N=4)}
\Description{A table of teacher demographics for the formative study (N=4), with columns Teacher ID, Age, Gender, and Teaching Experience in years. Formative-T1: age 45, male, 25 years. Formative-T2: age 44, male, 3 years, plus 15 years in assessment question design. Formative-T3: age 48, male, 20 years. Formative-T4: age 37, female, 5 years.}
\label{tab:formative_teacher_demographics}
\end{table}

\subsection{Framework Development Details}\label{appendix:framework_development}
\begin{table}[H]
\resizebox{\columnwidth}{!}{%
\centering
\begin{tabular}{lccc}
\toprule
\textbf{Teacher ID} & \textbf{Age} & \textbf{Gender} & \textbf{Teaching Exp. (yrs)} \\
\midrule
Framework-T1 & 52 & Female & Mathematics education researcher, Ph.D. \\
Framework-T2 & 37 & Male & 10 \\
Framework-T3 & 36 & Female & 10 \\
Framework-T4 & 34 & Female & 11 \\
Framework-T5 & 46 & Male & 20 \\
Framework-T6 & 49 & Male & 30 \\
\bottomrule
\end{tabular}
}
\caption{Teacher demographics for framework development process (N=6)}
\Description{A table of teacher demographics for the framework development process (N=6), with columns Teacher ID, Age, Gender, and Teaching Experience in years. Framework-T1: age 52, female, mathematics education researcher with a Ph.D. Framework-T2: age 37, male, 10 years. Framework-T3: age 36, female, 10 years. Framework-T4: age 34, female, 11 years. Framework-T5: age 46, male, 20 years. Framework-T6: age 49, male, 30 years.}
\label{tab:framework_teachers}
\end{table}
Table~\ref{tab:framework_teachers} summarizes the demographics of teachers who participated in iterative framework development. 

\subsection{Expert Review Details}\label{appendix:expert_review}
Table~\ref{tab:tech_eval_problems} presents the distribution of math word problems used in the expert review across grades, topics, and difficulty levels.
Table~\ref{tab:expert_review_participants} presents the demographics of all eight participants in the expert review, including six primary reviewers and two tiebreakers.
\begin{table}[H]
\resizebox{\columnwidth}{!}{%
\centering
\begin{tabular}{llccc}
\toprule
\textbf{Grade} & \textbf{Mathematical content domain} & \textbf{High} & \textbf{Medium} & \textbf{Total} \\
\midrule
\multirow{2}{*}{Grade 3} & Number \& Operations & 2 & 1 & 3 \\
& Geometry \& Measurement & 0 & 4 & 4 \\
\midrule
\multirow{2}{*}{Grade 4} & Number \& Operations & 0 & 3 & 3 \\
& Geometry \& Measurement & 0 & 2 & 2 \\
\midrule
\multirow{4}{*}{Grade 5} & Number \& Operations & 4 & 2 & 6 \\
& Geometry \& Measurement & 0 & 3 & 3 \\
& Change \& Measurement & 0 & 1 & 1 \\
& Data \& Probability & 1 & 1 & 2 \\
\midrule
\multirow{4}{*}{Grade 6} & Number \& Operations & 3 & 1 & 4 \\
& Geometry \& Measurement & 4 & 0 & 4 \\
& Change \& Measurement & 2 & 1 & 3 \\
& Data \& Probability & 1 & 0 & 1 \\
\midrule
\textbf{Total} & & \textbf{17} & \textbf{19} & \textbf{36} \\
\bottomrule
\end{tabular}
}
\caption{Distribution of math word problems used in the expert review across grades, mathematical content domains, and difficulty levels, where High and Medium refer to problem difficulty.}
\Description{A table showing the distribution of 36 math word problems used in the expert review across grades, mathematical content domains, and difficulty levels, with High and Medium referring to problem difficulty. Grade 3: Number and Operations, 2 high and 1 medium, total 3; Geometry and Measurement, 0 high and 4 medium, total 4. Grade 4: Number and Operations, 0 high and 3 medium, total 3; Geometry and Measurement, 0 high and 2 medium, total 2. Grade 5: Number and Operations, 4 high and 2 medium, total 6; Geometry and Measurement, 0 high and 3 medium, total 3; Change and Measurement, 0 high and 1 medium, total 1; Data and Probability, 1 high and 1 medium, total 2. Grade 6: Number and Operations, 3 high and 1 medium, total 4; Geometry and Measurement, 4 high and 0 medium, total 4; Change and Measurement, 2 high and 1 medium, total 3; Data and Probability, 1 high and 0 medium, total 1. Overall totals across all grades: 17 high, 19 medium, 36 problems.}
\label{tab:tech_eval_problems}
\end{table}
\begin{table}[H]
\resizebox{\columnwidth}{!}{%
\centering
\begin{tabular}{ccccc}
\toprule
\textbf{Teacher ID} & \textbf{Assigned Grade} & \textbf{Gender} & \textbf{Age} & \textbf{Teaching Experience (yrs)} \\
\midrule
E1 & Grade 6 & Female & 44 & 15 \\
E2 & Grade 6 & Female & 37 & 9.6 \\
E3 & Grade 5 & Male & 37 & 12 \\
E4 & Grade 5 & Female & 38 & 12 \\
E5 & Grades 3--4 & Male & 39 & 14 \\
E6 & Grades 3--4 & Male & 34 & 9 \\
E7 & Tiebreaker & Female & 39 & 5 \\
E8 & Tiebreaker & Male & 46 & 18 \\
\bottomrule
\end{tabular}
}
\caption{Teacher demographics for the expert review (N=8). E1--E6 were primary reviewers, each assigned to evaluate problems from a specific grade band. E7--E8 served as tiebreakers, resolving disagreements between primary reviewers.}
\Description{A table of teacher demographics for the expert review (N=8), where E1 through E6 were primary reviewers each assigned to a specific grade band, and E7 through E8 served as tiebreakers resolving disagreements between primary reviewers. E1: Grade 6, female, age 44, 15 years teaching experience. E2: Grade 6, female, age 37, 9.6 years. E3: Grade 5, male, age 37, 12 years. E4: Grade 5, female, age 38, 12 years. E5: Grades 3 to 4, male, age 39, 14 years. E6: Grades 3 to 4, male, age 34, 9 years. E7: Tiebreaker, female, age 39, 5 years. E8: Tiebreaker, male, age 46, 18 years.}
\label{tab:expert_review_participants}
\end{table}

\subsubsection{Evaluation Questions for Expert Review}\label{appendix:review_questions}
The evaluation questions used during expert review are listed below.

\noindent
\textbf{Sub-skill Level (Yes/No)}
\begin{itemize}
    \item \textbf{Construct Validity:} Do this sub-question and rubric appropriately assess the targeted problem-solving skill at the corresponding stage?
    \item \textbf{Level Appropriateness:} Are the complexity, information load, and phrasing of this sub-question and rubric suitable for the target grade level?
    \item \textbf{Mathematical Accuracy:} Are the exemplary answer and rubric criteria to this sub-question mathematically correct?
\end{itemize}
\textbf{Problem Level (Yes/No)}
\begin{itemize}
    \item \textbf{Problem-Context Alignment:} Does the overall set of sub-questions and rubrics naturally connect to the given problem context and effectively guide students through the problem-solving process?
    \item \textbf{Comprehensiveness:} Does the overall set of sub-questions and rubrics sufficiently cover the full problem-solving process for the given problem?
\end{itemize}

\subsection{Classroom Study Details}\label{appendix:classroom}
\subsubsection{Student Demographics}
Table~\ref{tab:classroom_student_grades} shows the grade distribution of the 64 students who participated in the classroom study.

\begin{table}[htbp]
\centering
\begin{tabular}{lc}
\toprule
\textbf{Grade} & \textbf{N} \\
\midrule
3 & 11 \\
4 & 16 \\
5 & 12 \\
6 & 14 \\
7 & 4 \\
8 & 5 \\
9 & 2 \\
\midrule
\textbf{Total} & \textbf{64} \\
\bottomrule
\end{tabular}
\caption{Grade distribution of participating students (N=64).}
\Description{A table showing the grade distribution of the 64 participating students. Grade 3: 11 students. Grade 4: 16 students. Grade 5: 12 students. Grade 6: 14 students. Grade 7: 4 students. Grade 8: 5 students. Grade 9: 2 students. Total: 64 students.}
\label{tab:classroom_student_grades}
\end{table}

\subsubsection{Student Survey Questions for Classroom Study}\label{appendix:student_survey}
The survey questions we used for the student survey in the classroom study are listed below:
\begin{itemize}
    \item \textbf{Usefulness of the Stage-Based Approach}: When solving the problem using this stage-based approach (i.e., breaking a large problem into multiple stages), did this method help you solve the problem more effectively?
    \item \textbf{Ease of Expressing Reasoning Process}: When using this approach, did answering the stage-based questions help you express your thinking more clearly?
    \item \textbf{Self-Awareness of Strengths and Weaknesses}: While solving the problem using this method, did you become more aware of which parts you did well and which parts you found difficult?
    \item \textbf{Willingness for Future Use}: Do you think this stage-based approach will be helpful for your future math learning?
    \item \textbf{Open-Ended Question}: Compared to your usual way of solving math problems, how was solving the problem using this stage-based approach different? What aspects did you find helpful, and what aspects did you find challenging?
\end{itemize}

\subsubsection{Diagnostic Report}
\autoref{fig:report-sample} shows a sample diagnostic report generated by \sysname{}, which summarizes the student's problem-solving skills at each stage.

\begin{figure*}[p]
    \centering
    \begin{subfigure}[b]{\linewidth}
        \centering
        \includegraphics[width=\linewidth,height=0.85\textheight,keepaspectratio]{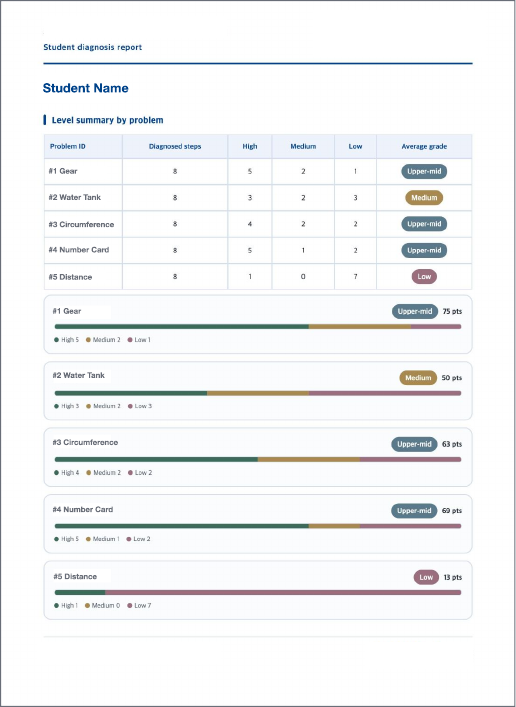}
        \caption{}
        \label{fig:report-sample-a}
    \end{subfigure}
\end{figure*}

\begin{figure*}[p]
    \ContinuedFloat
    \centering
    \begin{subfigure}[b]{\linewidth}
        \centering
        \includegraphics[width=\linewidth,height=0.85\textheight,keepaspectratio]{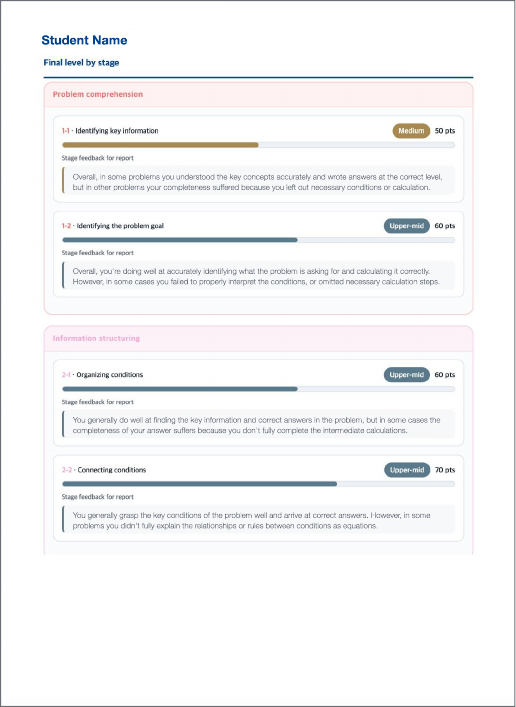}
        \caption{}
        \label{fig:report-sample-b}
    \end{subfigure}
\end{figure*}

\begin{figure*}[p]
    \ContinuedFloat
    \centering
    \begin{subfigure}[b]{\linewidth}
        \centering
        \includegraphics[width=\linewidth,height=0.85\textheight,keepaspectratio]{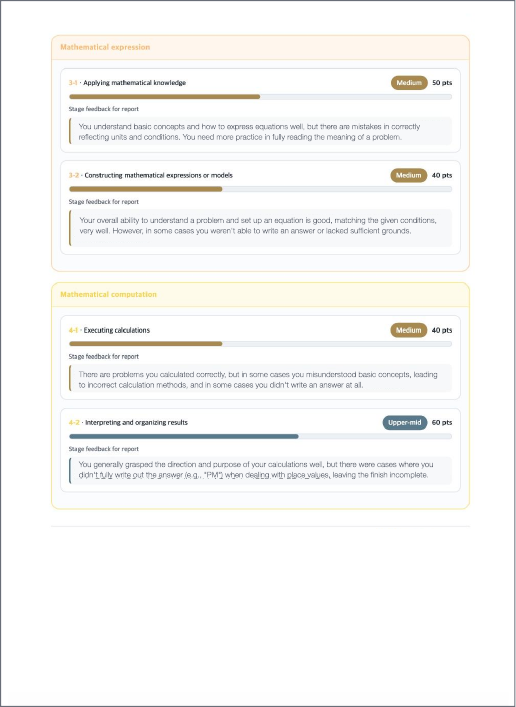}
        \caption{}
        \label{fig:report-sample-c}
    \end{subfigure}
    \caption{A sample student's diagnostic report generated by \sysname{}. (a) A summary view showing performance across all problems, with progress bars, High/Medium/Low score breakdowns, and an overall level badge for each problem. (b)–(c) A detailed, stage-by-stage view for a single problem, grouped by the STEP framework's four stages, where each sub-skill shows a progress bar, numerical score, and a text summary of the student's strengths and areas for improvement.}
    \Description{Three stacked screenshots of a student diagnostic report generated by STEPwise. The first screenshot (a) shows a summary view listing five math problems, each with a horizontal progress bar, a breakdown of High/Mid/Low counts, and an overall level badge ranging from Low to Upper-Mid. The second and third screenshots (b, c) show a detailed per-stage view for one problem, organized under the STEP framework's four stages: Problem Comprehension, Information Structuring, Mathematical Expression, and Mathematical Computation. Each sub-skill entry includes a progress bar, a numerical score, and a short text summary describing the student's strengths and weaknesses at that stage.}
    \label{fig:report-sample}
\end{figure*} 
% 1028
\end{document}